\documentclass[conference]{IEEEtran}
\IEEEoverridecommandlockouts

\usepackage{amsfonts}
\usepackage{textcomp}
\usepackage{xcolor}
\usepackage{fancyhdr}
\usepackage{cite}
\usepackage{graphicx}
\usepackage{psfrag}
\usepackage{subfigure}
\usepackage{url}
\usepackage{stfloats}
\usepackage{amsmath}
\usepackage{array}
\usepackage{fancyhdr}
\usepackage{epsfig}
\usepackage{amssymb}
\usepackage{color}
\usepackage{float}
\usepackage{algorithm}
\usepackage{algpseudocode}
\usepackage{balance}
\usepackage{soul}

\def\BibTeX{{\rm B\kern-.05em{\sc i\kern-.025em b}\kern-.08em
    T\kern-.1667em\lower.7ex\hbox{E}\kern-.125emX}}
\begin{document}

\title{Satellite-based Quantum Network: Security and Challenges over Atmospheric Channel}
\author{\IEEEauthorblockN{Hong-fu Chou, Vu Nguyen Ha, Hayder Al-Hraishawi, Luis Manuel Garces-Socarras, \\ Jorge Luis Gonzalez-Rios, Juan Carlos  Merlano-Duncan, Symeon Chatzinotas}
\IEEEauthorblockA{\textit{Interdisciplinary Centre for Security, Reliability and Trust (SnT), University of Luxembourg, Luxembourg}\\ 
Emails: \{hungpu.chou,vu-nguyen.ha,hayder.al-hraishaw,luis.garces,jorge.gonzalez,juan.duncan,symeon.chatzinotas\}@uni.lu}
}

\maketitle
\begin{abstract}
The ultra-secure quantum network leverages quantum cryptography to deliver unsurpassed data transfer security. In principle, the well-known quantum key distribution (QKD) achieves unconditional security, which raises concerns about the trustworthiness of 6G wireless systems in order to mitigate the gap between practice and theory. The long-distance satellite-to-ground evolving quantum network distributes keys that are ubiquitous to the node on the ground through low-orbit satellites. As the secret key sequence is encoded into quantum states, it is sent through the atmosphere via a quantum channel. It still requires more effort in the physical layer design of deployment ranges, transmission, and security to achieve high-quality quantum communication. In this paper, we first review the quantum states and channel properties for satellite-based quantum networks and long-range quantum state transfer (QST). Moreover, we highlight some challenges, such as transmissivity statistics, estimation of channel parameters and attack resilience, quantum state transfer for satellite-based quantum networks, and wavepacket shaping techniques over atmospheric channels. We underline two research directions that consider the QST and wavepacket shaping techniques for atmospheric transmission in order to encourage further research toward the next generation of satellite-based quantum networks. 
\end{abstract}
\begin{IEEEkeywords}
Quantum key distribution, satellite communication, quantum state transfer, wavepacket shaping modulation
\end{IEEEkeywords}

\section{Introduction}
To meet the growing global demand for broadband services, satellites offer the unique ability to cover enormous geographic areas while requiring minimum base infrastructure, making them an appealing alternative\cite{Symeon2019}. The ability to connect quantum devices across long distances, considerably boosting their inherent communication, network efficiency, and security, is made possible by an examination of the state-of-the-art of the key components of quantum networks\cite{Parny_2023}. By delivering quantum states containing critical information through free space optical (FSO) channels, the authors in \cite{Trinh2018} reveal the possibility of establishing global quantum networks. To achieve unconditional security, quantum key distribution (QKD)\cite{QKD1,QKD2} secures a protocol that can guarantee information security in theory between two remote nodes while exchanging keys. QKD is the most thoroughly explored quantum communication technology\cite{Lajos}, and it has been used in both fiber and FSO channels. Long-distances FSO quantum communications\cite{Peng_2005,Resch_2005,Erven_2008} have been deployed effectively across extraordinarily long distances while several experiments have also been realized in \cite{Schmit2007,Scheidl2009,Fedrizzi_2009}. However, when channel attenuation and noise levels increase in the communication via fiber or FSO, the communication range for successful key distribution is restricted to a few hundred kilometers. That raises a significant challenge in satellite-based QKD systems.
Recently, quantum teleportation\cite{Yin_2012,Ma_2012,Liao2017LongdistanceFQ,Tasio2023} has been considered as a possible approach for increasing the deployment range of QKD using satellites. By moving to free space and connecting two ground stations, the effective communication range can be increased to a hundred km; while connecting a ground station with a satellite, quantum teleportation was conducted across a thousand km. Although channel noise has no direct effect on the transported qubit, it does result in an entanglement network with low fidelity, diminishing the success chance of secret key transmission by quantum teleportation\cite{Zhonghui2023}. Furthermore, while QKD theoretically provides ultra-security to satellite-based QKD networks, satellite-to-ground, and inter-satellite quantum communication systems, the exploitation of defects in quantum cryptography protocols to undermine information security or obtain unauthorized access to sensitive data is referred to as quantum hacking\cite{Zhao_2008,Scarani2009,Diamanti_2016}. The physical layer security deduces that the efficiency for detecting binary bit levels is identical, this raises doubts about the validity of the security proof. The time-shift attack makes use of a QKD system's time-domain detection efficiency discrepancy between its two detectors. It is only possible to securely eliminate the time-domain inconsistency of detection efficiency. Therefore, the characteristics of the atmosphere can have identical security concerns and are the critical instance required to consider and investigate.

Starting with a background study of atmospheric channel characteristics carrying quantum states and progressing to research challenging issues, the goal of this paper is to bring up some more in-depth investigation by reviewing and proposing open challenges of physical layer transmission and security via satellite-based QKD networks, which will necessitate developing more secure and efficient scenarios to achieve high-quality quantum communications. We hope to explore the following significant implementation issues as a result of this:
\begin{itemize}
  \item It is critical to investigate system design for the satellite-based quantum networks over atmospheric channels in order to assure optimal system performance, as well as approaches for diversified scheduling for space missions or satellite-to-ground communication.
  \item Because interference issues caused by coexistence with other satellite systems, as well as atmospheric turbulence and diffraction, affect the fidelity of satellite-based quantum networks, developing novel strategic deployment and interference mitigation techniques is critical.
  \item By using quantum mechanical ideas in QKD networks, the quantum states based on a variety of QKD protocols, as well as the heterogeneity of transmissivity from satellite-to-ground/air/space-destination scenarios to the aforementioned deployments, can be realized. As a result, channel parameter estimation is indispensable in such dynamic satellite communication.
  \item Experiments show that the assaults against a commercial QKD system are technically feasible. Because of these minor flaws, satellite-based QKD networks may be susceptible to Eve with current technology. The effectiveness of the attacks emphasizes the need to develop innovative techniques for security proofs with verifiable premises and battle-test actual resilience to any attacks. 
\end{itemize}
Additional prospects for satellite-based quantum networks and potential future research topics are also included in this paper.
\section{Quantum networks via Atmospheric Channel}
We start from a basic knowledge of quantum states in QKD networks and atmospheric channels to the potential future research avenues.
\subsection{Entangled Gaussian and Non-Gaussian States}
Continuous variables Gaussian states, such as thermal, coherent, and compressed states of light, are one example of states that cannot be purified by ordinary processes\cite{Rod__2007}. When ordinary processes are referred to, the actions that keep the state's Gaussian characteristics intact are implicit in matching mirrors, beam splitters, and squeezers. Therefore, any entangled Gaussian state has bound entanglement in the Gaussian assumption. The symplectic matrix allows for a compact expression of the canonical commutation relations (CCR)\cite{Rod__2007} for a quantum system with n modes. Gaussian two-mode squeezed vacuum (TMSV) states are characterized by a gaussian $\chi _{\rho }$ function.
\begin{equation}
\chi _{\rho }(\zeta )=e^{i\zeta ^{T}J_{n}d-\frac{1}{4}\zeta^{T}J_{T}^{n}\gamma J_{n }\zeta }
\end{equation}
where d is a $2n$ real vector, called displacement vector (DV), and $\gamma$ is a $2n \times 2n$
symmetric real matrix, denoted as the covariance matrix (CM).
Researchers\cite{Grosshans_2002} apply the scenario of prepare and measure with compressed or coherent states to experimentally build quantum cryptography with gaussian states and gaussian operations to discuss the efficiency of security protocol. 

Implementing non-Gaussian entangled states effectively across atmospheric fading channels is still difficult while maintaining a reliable secret key rate.
Recent research has looked at the rate of entanglement formation caused by non-Gaussian entangled states moving through atmospheric channels \cite{Hosseinidehaj_2015,Hosseinidehaj_2016}. A non-Gaussian
entangled state \cite{Hosseinidehaj_2016} can be addressed by giving a Gaussian state and zero first moment CM
by
\begin{equation}
    M_{AB}^{in} = \begin{bmatrix}
        vI & \sqrt{v^{2}-1}Z \\
        \sqrt{v^{2}-1}Z & vI \\   
    \end{bmatrix}
\end{equation}
where $I$ is a $2 \times  2$ identity matrix, Z is a $(1, -1)$ diagonal matrix, and
$v = cosh (2r)$ is the quadrature variance of each mode. The pure photon-subtracted squeezed (PSS) state is introduced to the unmeasured output of the beam splitter in order to elicit this operation of non-Gaussian states by detecting heralded modes. This transfer is then analyzed by using the Kraus representation to chart the PSS state's progress through the fading channel. The authors\cite{Hosseinidehaj_2016} investigate that contrary to common sense, non-Gaussian states can occasionally yield greater quantum key rates via fixed-attenuation channels and particularly for very high-loss channels. Furthermore, the measurement result\cite{Liao_2018} is determined using a coherent detector at the receiver using an upgraded non-Gaussian state discrimination detector. The received nonorthogonal coherent states are measured with the state-discrimination detector, which is regarded as the optimal quantum measurement\cite{Liao_2018}. High uplink losses are frequently an issue for satellite-based communication systems. Without the helpful intervention, satellite-based entanglement distribution and quantum key distribution would remain an unsuccessful undertaking. Therefore, we discuss the atmospheric channel loss in the next section.
\subsection{Channel Loss of Satellite-based quantum networks }
For the satellite-based QKD networks, the uplink and downlink channels are quite different. On an uplink channel, the atmospheric turbulence layer only occurs close to the transmitter, while on a downlink, it only exists close to the terrestrial receiver. The uplink optical beam initially travels through the turbulent environment for typical ground station aperture sizes, and its beam-width is substantially smaller than the large-scale turbulent vortex. The downlink optical beam only passes through the turbulent environment during the last portion of its route, in contrast to the uplink channel. The satellite's beam-width upon entrance into the atmosphere is most likely to be greater than the size of the turbulent vortex given the usual aperture size of the optical equipment incorporated in the satellite. Therefore, turbulence-induced effects \cite{Lajos} cause the transmissivity, $\eta _{t}$, of atmospheric channels to change. The probability distribution of the transmission coefficients, indicated by $p(\eta)$, may be used to describe these fading channels where  $\eta = \sqrt{\eta _{t}}$. The mean fading loss in dB for a channel that is fading and associated with the probability distribution $p(\eta)$ is given by $-10log_{10}\int _{0}^{\eta _{0}}\eta ^{2}p(\eta )d\eta $, where $\eta _{0}$ is the greatest value of $\eta$. Another dominant of the loss is beam-wandering which causes the beam-center to wander erratically from the receiver's aperture plane, regardless of atmospheric turbulence-related changes in beam-width\cite{Lajos}. 

When beam-width variations are taken into account, the mean fading loss in dB of a fading channel is now given by $-10log_{10}\int \eta ^{2}(l,\theta )p(l,\theta  )dld\theta  $ with the knowledge in \cite{Lajos}. It should be noted that, with the addition of beam-width variations, the channel's maximum transmission coefficient $\eta _{0}$ is no longer constant but instead varies at random. Therefore, the above introduction can lead us to the research in \cite{Bohmann_2016} to investigate Gaussian entangled states in fading channels and uncorrelated fading channels. The transmitted continuous-variable entanglement has a non-trivial effect on coherent displacements of the input field's quantum state in such turbulent channels. Surprisingly, this enables one to maximize the entanglement certification by altering local coherent amplitudes with a limited but optimal amount of squeezing. These variables can be changed to enhance the Gaussian entanglement transfer via a turbulent environment by adaptive method attaining the correlated form in the entanglement preserving link. Following this thread, the long-range quantum state transfer is presented in the following section. 
\subsection{Long-distance Quantum State Transfer}
Many quantum information processing (QIP)\cite{Tony2022} activities rely on the transport of quantum information between various sites. The quantum state transfer (QST)\cite{Huang_2021} of interacting qubits might be significantly more advantageous, both for the QIP protocol and for transmitting the entire physical configuration of quantum communication.
Long-distance QST is an important component of quantum protocols that can be realized by quantum teleportation. The challenges of long-distance QST\cite{Almeida_2018,Hermes_2020,Shen_2023} are to maintain the fidelity and reduce the low fidelity caused by inconsistency of the channel's mirror symmetry. Moreover, in prior long-distance investigations, Alice and Charlie have always performed local Bell-state measurements before the entanglement distribution procedure. It is difficult to conduct the Bell-state measurement after photons pass via air channels due to atmospheric turbulence. The authors in \cite{Bo2022} prove the QST over a distance of more than 1000 km with the satellite Micius aided by previous quantum entanglement shared between two distant ground stations. The highly steady interferometer project the photon into a composite path-polarization dimension and makes use of a satellite-borne entangled photon source. 
\section{Open Challenges}
\begin{figure*}[htb]
\begin{center}
\includegraphics[trim={0cm 0cm 0cm 0cm},width= 95mm,height = 41mm]{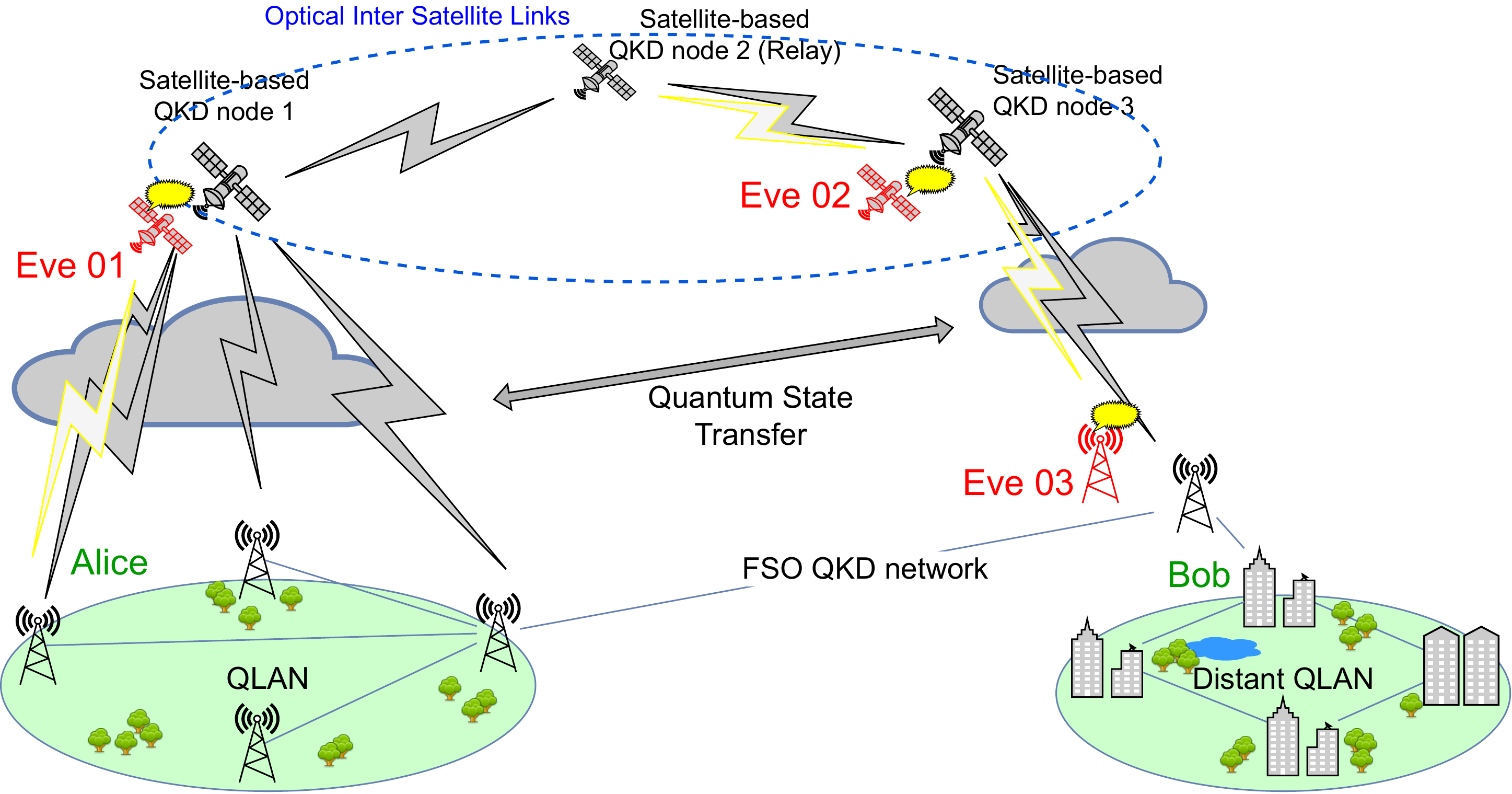} 
\caption{A Vision of Satellite-based Quantum Networks.}
\label{QST}
\end{center}
\end{figure*}

Based on the extensive approaches for making quantum states adapt to atmospheric channel variations, we identify the following challenges in order to increase fidelity and security resilience, more research efforts should be addressed for greater robustness to actual attacks, as well as the inspiration for a better understanding of quantum state correlation in the atmosphere. We depict the quantum networks in Fig. \ref{QST} to illustrate the following challenges.
\subsection{Transmissivity Statistics on Atmospheric Channels}
Extensive theoretical examinations\cite{Bohmann_2017,Vasylyev_2018,Hofmann_2019} of these effects are presented, and these models are shown to be broadly coherent with terrestrial testing done under a broad range of turbulence conditions.
The quantum states of entanglement conditions were explored in \cite{Bohmann_2017,Eduardo2021,Zhao2022MonteCS} based on the negativity of the partial transposition. In order to implement and optimize quantum communication in atmospheric connections, the partial transposition can be viewed as the output state of turbulent atmospheric fluctuating-loss channels. Despite the dominating impacts of the variable nature of transmissivity in these channels, which introduces extra noise and limits the possible secret key rate, other more subtle atmospheric effects such as weather condition\cite{Heim_2009,Vasylyev_2018} and day-and-night times\cite{Liao2017LongdistanceFQ}, which consider beam attenuation, absorption, and scattering, can play a significant role\cite{Nedasada2017,Hosseinidehaj_2016,Hosseinidehaj_2015,Keshav2023}. We illustrate in Fig. \ref{QST} the case of satellite-based QKD node 1 communicating with the ground user terminal and long-range communication between satellite-based QKD node 1, 2, and 3. The collaboration of satellite-based QKD node 2 and 3 communicates to the ground user terminal in distance QLAN. Due to the above discussion, QKD networks are very sensitive to the channel’s
transmissivity even during day-and-night times. A thorough understanding of atmospheric channels\cite{Wang_2020,Liao2017LongdistanceFQ,Khmelev_2021} has broad implications for research, and utilizing satellite-based nodes and quantum ground transceivers with the bulk of propagation occurring outside the Earth's atmosphere can enable global-scale quantum communication. Furthermore, transmissivity and secrete key rate can be aided by self-compensating technique\cite{Erven_2008,Agnesi_2020,Osborn2021} such as some use cases of passive polarization analysis units, adaptive optics pre-compensation,
GPS timing receivers for synchronization, and custom-written software. This research trend is still in the initiative stage, and thus, more sophisticated approaches are necessary to optimize the satellite-based QKD networks. Additionally, it is noteworthy to remark that the authors in \cite{wang2022exploiting} present an interesting framework for monitoring the continually changing satellite-to-ground atmospheric channel. This framework aspires to enhance the scheduling techniques by optimizing the available time for SatQKD and reducing the average link loss during a single orbit. The suggested scheduling technique serves the dual objective of reducing geometric loss in future satellite orbit selection and enhancing the effective use of designated keys. 
\subsection{Estimation of Channel Parameters and Attack Resilience}
The general idea of the design requirement for satellite-based QKD networks is that because of the changing nature of transmissivity in these channels, more noise is generated, limiting the feasible secret key rate. A thorough framework for analyzing the needs for spatial, spectral\cite{Chin2019}, and temporal filtering is provided by the study\cite{Lanning2021,Zheng2023} for wavelength-dependence free space QKD networks. However, an adaptable paradigm for optimal wavelength is required more effort to provide high performance. Despite the practical attack on QKD networks\cite{Zhao_2008}, channel parameter estimation is not only beneficial to secret key rate performance\cite{Takenaka_2017 ,Ying2018,Dequal_2021,Peng_2022} but also security concern \cite{Chai2019,Chai2020BlindCE} to demonstrate against entanglement-distillation attack. Eve may be undetected by Alice and Bob utilizing basic light monitoring that is present in quantum repeaters such as satellite-to-ground channels and ground-to-ground networks in the event of this assault potentially generating a loophole based on the assumption \cite{Guo2017}. The Monte Carlo approach\cite{Chai2019} was used to estimate the channel transmittance, which changes randomly over time. Especially when there is little knowledge of the channel, as there is in the case of the atmosphere and the ocean, it is improper for parametric variation channels to be based on insufficient statistical features. Authorized communication\cite{Liao_2018} is required to continuously communicate partial input and output signals in order to update the channel estimate when the channel changes, even if it changes slowly. As illustrated in Fig. \ref{QST}, both Eve 1 and 2 present the threat to demonstrate the potential loophole between Alice and Bob. Moreover, malicious attackers in the ground user terminal act as Eve 3 performing unauthorized access to QKD node 3. Above all, further research efforts must be made in order to investigate both the secret key rate performance and the durability of the attack.   
\subsection{The Quantum State Transfer for Satellite-based Quantum Networks}
Following the discussion in Section II D, QST allows the transfer of arbitrary quantum states from one node to another by SWAP\cite{Lu_2019} or quantum teleportation. As illustrated in Fig. \ref{QST}, the quantum state transfer exist over the atmosphere for the satellite-based QKD networks. Moreover, spatial steering is well-known that quantum states may be remotely created via an entangled pair to be benchmarked by average fidelity\cite{Huang_2021}. It is worth noting that the study of a quantum teleportation protocol in \cite{Tasio2023} presents a comprehensive investigation of uplink and downlink channel loss mechanisms. This thread leads us to the challenge of investigating the effectiveness of QST on atmospheric channels. The authors in \cite{Pande_2022} first touched on this challenge. Regarding remote state determination or remote state tomography,  quantum information transfer considers providing a quantum teleportation protocol and associated security discussion with the QKD protocol by means of high-dimension quantum correlations. Furthermore, the experiment results of quantum teleportation \cite{Tasio2023} show improved fidelity by the generation of states in an intermediate station that produces nearly symmetric states in line with the well-known Braunstein-Kimble quantum teleportation process to have a better teleportation fidelity. Furthermore, the authors in \cite{Semenov_2010} provide the experiment of a given event proving that the concurrence of entanglement can be conserved by turbulent atmospheric channel. By demonstrating how to improve the fidelity of QST or entangled transfer on atmosphere measuring prior to the privacy amplification of satellite-based QKD networks, we highlight this potential research avenue that can be taken into consideration in the physical layer of quantum networks. In order to improve the secret key rate, transmissivity, and fidelity for sophisticated satellite-based quantum networks, it is beneficial to investigate the quantum state correlation and decoherence on atmospheric channels, which depend on the implementation of entanglement transfer or QST.

\subsection{Wavepacket Shaping Techniques over Atmospheric Channels}
Following the quantum state introduction of Section II A, an innovation technique in \cite{Chuu2021,Wu_2019} observes the purification and entanglement after conducting wavepacket shaping modulation to single photon and entangled photon which is applicable to both single-photon and entangled-biphoton QKD devices. The experiments in \cite{Chuu2021} demonstrates how the wavepacket modulation control photons emitted from the colloidal quantum dots at room temperature. It has been shown that biexciton emission is eliminated by modulating the wavepacket, the emitted single photons preserve the high purity even at high excitation power. Furthermore, the result of concurrence and purity shows that wavepacket modulation reserve and restore the entanglement at the high-frequency difference by topographically reconstructing the density matrix. However, this shaping modulation applies to satellite-based quantum networks which remain more explored and studied over atmospheric channels. Based on the purification and revival entanglement to the quantum state, the implicit correlation with long-range QST to the physical layer of the quantum network is still not clear about how the photon absorption efficiency and revival of entanglement can enhance the transmissivity and entangled quantum networks respectively. This could be quite an interesting research direction to investigate for the advanced design of physical layers in satellite-based quantum networks.

\section{Conclusion and Future Directions}
In order to enhance the quality of long-range quantum communication, this paper reviews several significant technical evaluations of satellite-based quantum networks and suggests future research options. First, the knowledge of quantum states provokes the understanding of a fundamental transmission unit for quantum communication. The atmospheric channel loss is explored to depict the effect on coherent displacements and entanglement certification of the quantum state. A more sophisticated self-compensating approach to receiver design for satellite-based quantum networks can help to improve transmissivity and secret key rate. The channel estimate is therefore seen as a challenge in terms of both ensuring higher key transmission quality and thwarting possible attacks. Based on the above reviews, this paper features two open challenges that the implementation of QST and wavepacket shaping technique are required to investigate how to improve the security, fidelity, and transmissivity of satellite-based quantum networks and can be the potential candidates for improving the existing approach to the challenges of estimating channel parameters and attack resilience. 

\section*{Acknowledgment}
This work is under the project LUQCIA funded by the European Union Next Generation EU, with the collaboration of the Department of Media, Connectivity, and Digital Policy (SMC), Luxembourg.

\bibliographystyle{IEEEtran}
\bibliography{gcomwsqkd}

\begin{thebibliography}{10}
\providecommand{\url}[1]{#1}
\csname url@samestyle\endcsname
\providecommand{\newblock}{\relax}
\providecommand{\bibinfo}[2]{#2}
\providecommand{\BIBentrySTDinterwordspacing}{\spaceskip=0pt\relax}
\providecommand{\BIBentryALTinterwordstretchfactor}{4}
\providecommand{\BIBentryALTinterwordspacing}{\spaceskip=\fontdimen2\font plus
\BIBentryALTinterwordstretchfactor\fontdimen3\font minus
  \fontdimen4\font\relax}
\providecommand{\BIBforeignlanguage}[2]{{%
\expandafter\ifx\csname l@#1\endcsname\relax
\typeout{** WARNING: IEEEtran.bst: No hyphenation pattern has been}%
\typeout{** loaded for the language `#1'. Using the pattern for}%
\typeout{** the default language instead.}%
\else
\language=\csname l@#1\endcsname
\fi
#2}}
\providecommand{\BIBdecl}{\relax}
\BIBdecl

\bibitem{Symeon2019}
A.~I. Perez-Neira, M.~A. Vazquez, M.~B. Shankar, S.~Maleki, and S.~Chatzinotas,
  ``Signal processing for high-throughput satellites: Challenges in new
  interference-limited scenarios,'' \emph{IEEE Signal Processing Magazine},
  vol.~36, no.~4, pp. 112--131, 2019.

\bibitem{Parny_2023}
\BIBentryALTinterwordspacing
L.~de~Forges~de Parny, O.~Alibart, J.~Debaud, S.~Gressani, A.~Lagarrigue,
  A.~Martin, A.~Metrat, M.~Schiavon, T.~Troisi, E.~Diamanti, P.~G{\'{e}}lard,
  E.~Kerstel, S.~Tanzilli, and M.~V.~D. Bossche, ``Satellite-based quantum
  information networks: use cases, architecture, and roadmap,''
  \emph{Communications Physics}, vol.~6, no.~1, jan 2023. [Online]. Available:
  \url{https://doi.org/10.1038%2Fs42005-022-01123-7}
\BIBentrySTDinterwordspacing

\bibitem{Trinh2018}
P.~V. Trinh, A.~T. Pham, A.~Carrasco-Casado, and M.~Toyoshima, ``Quantum key
  distribution over fso: Current development and future perspectives,'' in
  \emph{2018 Progress in Electromagnetics Research Symposium (PIERS-Toyama)},
  2018, pp. 1672--1679.

\bibitem{QKD1}
\BIBentryALTinterwordspacing
N.~Gisin, G.~Ribordy, W.~Tittel, and H.~Zbinden, ``Quantum cryptography,''
  \emph{Rev. Mod. Phys.}, vol.~74, pp. 145--195, Mar 2002. [Online]. Available:
  \url{https://link.aps.org/doi/10.1103/RevModPhys.74.145}
\BIBentrySTDinterwordspacing

\bibitem{QKD2}
D.~Mayers, ``Unconditional security in quantum cryptography,'' 2004.

\bibitem{Lajos}
N.~Hosseinidehaj, Z.~Babar, R.~Malaney, S.~X. Ng, and L.~Hanzo,
  ``Satellite-based continuous-variable quantum communications:
  State-of-the-art and a predictive outlook,'' \emph{IEEE Communications
  Surveys \& Tutorials}, vol.~21, no.~1, pp. 881--919, 2019.

\bibitem{Peng_2005}
\BIBentryALTinterwordspacing
C.-Z. Peng, T.~Yang, X.-H. Bao, J.~Zhang, X.-M. Jin, F.-Y. Feng, B.~Yang,
  J.~Yang, J.~Yin, Q.~Zhang, N.~Li, B.-L. Tian, and J.-W. Pan, ``Experimental
  free-space distribution of entangled photon pairs over 13~km: Towards
  satellite-based global quantum communication,'' \emph{Physical Review
  Letters}, vol.~94, no.~15, apr 2005. [Online]. Available:
  \url{https://doi.org/10.1103%2Fphysrevlett.94.150501}
\BIBentrySTDinterwordspacing

\bibitem{Resch_2005}
K.~Resch, M.~Lindenthal, B.~Blauensteiner, H.~Bohm, A.~Fedrizzi, M.~Taraba,
  R.~Ursin, P.~Walther, A.~Poppe, T.~Schmitt-Manderbach, H.~Weier,
  H.~Weinfurter, and A.~Zeilinger, ``Distributing entanglement and single
  photons through an intra-city, free-space quantum channel,'' in \emph{EQEC
  '05. European Quantum Electronics Conference, 2005.}, 2005, pp. 298--.

\bibitem{Erven_2008}
\BIBentryALTinterwordspacing
C.~Erven, C.~Couteau, R.~Laflamme, and G.~Weihs, ``Entangled quantum key
  distribution over two free-space optical links,'' \emph{Optics Express},
  vol.~16, no.~21, p. 16840, oct 2008. [Online]. Available:
  \url{https://doi.org/10.1364%2Foe.16.016840}
\BIBentrySTDinterwordspacing

\bibitem{Schmit2007}
\BIBentryALTinterwordspacing
T.~Schmitt-Manderbach, H.~Weier, M.~F\"urst, R.~Ursin, F.~Tiefenbacher,
  T.~Scheidl, J.~Perdigues, Z.~Sodnik, C.~Kurtsiefer, J.~G. Rarity,
  A.~Zeilinger, and H.~Weinfurter, ``Experimental demonstration of free-space
  decoy-state quantum key distribution over 144 km,'' \emph{Phys. Rev. Lett.},
  vol.~98, p. 010504, Jan 2007. [Online]. Available:
  \url{https://link.aps.org/doi/10.1103/PhysRevLett.98.010504}
\BIBentrySTDinterwordspacing

\bibitem{Scheidl2009}
T.~Scheidl, R.~Ursin, A.~Fedrizzi, S.~Ramelow, X.~song Ma, T.~Herbst,
  R.~Prevedel, L.~Ratschbacher, J.~Kofler, T.~Jennewein, and A.~Zeilinger,
  ``Feasibility of 300km quantum key distribution with entangled states,''
  \emph{New Journal of Physics}, vol.~11, p. 085002, 2009.

\bibitem{Fedrizzi_2009}
\BIBentryALTinterwordspacing
A.~Fedrizzi, R.~Ursin, T.~Herbst, M.~Nespoli, R.~Prevedel, T.~Scheidl,
  F.~Tiefenbacher, T.~Jennewein, and A.~Zeilinger, ``High-fidelity transmission
  of entanglement over a high-loss free-space channel,'' \emph{Nature Physics},
  vol.~5, no.~6, pp. 389--392, may 2009. [Online]. Available:
  \url{https://doi.org/10.1038%2Fnphys1255}
\BIBentrySTDinterwordspacing

\bibitem{Yin_2012}
\BIBentryALTinterwordspacing
J.~Yin, J.-G. Ren, H.~Lu, Y.~Cao, H.-L. Yong, Y.-P. Wu, C.~Liu, S.-K. Liao,
  F.~Zhou, Y.~Jiang, X.-D. Cai, P.~Xu, G.-S. Pan, J.-J. Jia, Y.-M. Huang,
  H.~Yin, J.-Y. Wang, Y.-A. Chen, C.-Z. Peng, and J.-W. Pan, ``Quantum
  teleportation and entanglement distribution over 100-kilometre free-space
  channels,'' \emph{Nature}, vol. 488, no. 7410, pp. 185--188, aug 2012.
  [Online]. Available: \url{https://doi.org/10.1038%2Fnature11332}
\BIBentrySTDinterwordspacing

\bibitem{Ma_2012}
\BIBentryALTinterwordspacing
X.-S. Ma, T.~Herbst, T.~Scheidl, D.~Wang, S.~Kropatschek, W.~Naylor,
  B.~Wittmann, A.~Mech, J.~Kofler, E.~Anisimova, V.~Makarov, T.~Jennewein,
  R.~Ursin, and A.~Zeilinger, ``Quantum teleportation over 143 kilometres using
  active feed-forward,'' \emph{Nature}, pp. 269--273, sep 2012. [Online].
  Available: \url{https://doi.org/10.1038%2Fnature11472}
\BIBentrySTDinterwordspacing

\bibitem{Liao2017LongdistanceFQ}
S.~Liao, H.-L. Yong, C.~Y. Liu, G.~Shentu, D.-D. Li, J.~Lin, H.~Dai, S.-Q.
  Zhao, B.~Li, J.-Y. Guan, W.~Chen, Y.-H. Gong, Y.~Li, Z.~Lin, G.-S. Pan, J.~S.
  Pelc, M.~M. Fejer, W.~zhuo Zhang, W.~Liu, J.~Yin, J.-G. Ren, X.-B. Wang,
  Q.~Zhang, C.-Z. Peng, and J.-W. Pan, ``Long-distance free-space quantum key
  distribution in daylight towards inter-satellite communication,''
  \emph{Nature Photonics}, vol.~11, pp. 509 -- 513, 2017.

\bibitem{Tasio2023}
T.~Gonzalez-Raya, S.~Pirandola, and M.~Sanz, ``Satellite-based entanglement
  distribution and quantum teleportation with continuous variables,'' 2023.

\bibitem{Zhonghui2023}
Z.~Li, K.~Xue, J.~Li, L.~Chen, R.~Li, Z.~Wang, N.~Yu, D.~S. Wei, Q.~Sun, and
  J.~Lu, ``Entanglement-assisted quantum networks: Mechanics, enabling
  technologies, challenges, and research directions,'' \emph{IEEE
  Communications Surveys \& Tutorials}, pp. 1--1, 2023.

\bibitem{Zhao_2008}
\BIBentryALTinterwordspacing
Y.~Zhao, C.-H.~F. Fung, B.~Qi, C.~Chen, and H.-K. Lo, ``Quantum hacking:
  Experimental demonstration of time-shift attack against practical
  quantum-key-distribution systems,'' \emph{Physical Review A}, vol.~78, no.~4,
  oct 2008. [Online]. Available:
  \url{https://doi.org/10.1103%2Fphysreva.78.042333}
\BIBentrySTDinterwordspacing

\bibitem{Scarani2009}
\BIBentryALTinterwordspacing
V.~Scarani, H.~Bechmann-Pasquinucci, N.~J. Cerf, M.~Du\ifmmode~\check{s}\else
  \v{s}\fi{}ek, N.~L\"utkenhaus, and M.~Peev, ``The security of practical
  quantum key distribution,'' \emph{Rev. Mod. Phys.}, vol.~81, pp. 1301--1350,
  Sep 2009. [Online]. Available:
  \url{https://link.aps.org/doi/10.1103/RevModPhys.81.1301}
\BIBentrySTDinterwordspacing

\bibitem{Diamanti_2016}
\BIBentryALTinterwordspacing
E.~Diamanti, H.-K. Lo, B.~Qi, and Z.~Yuan, ``Practical challenges in quantum
  key distribution,'' \emph{npj Quantum Information}, vol.~2, no.~1, nov 2016.
  [Online]. Available: \url{https://doi.org/10.1038%2Fnpjqi.2016.25}
\BIBentrySTDinterwordspacing

\bibitem{Rod__2007}
\BIBentryALTinterwordspacing
C.~Rod{\'{o} }, O.~Romero-Isart, K.~Eckert, and A.~Sanpera, ``Efficiency in
  quantum key distribution protocols with entangled gaussian states,''
  \emph{Open Systems Information Dynamics}, vol.~14, no.~01, pp. 69--80, mar
  2007. [Online]. Available: \url{https://doi.org/10.1007%2Fs11080-007-9030-x}
\BIBentrySTDinterwordspacing

\bibitem{Grosshans_2002}
\BIBentryALTinterwordspacing
F.~Grosshans and P.~Grangier, ``Continuous variable quantum cryptography using
  coherent states,'' \emph{Physical Review Letters}, vol.~88, no.~5, jan 2002.
  [Online]. Available: \url{https://doi.org/10.1103%2Fphysrevlett.88.057902}
\BIBentrySTDinterwordspacing

\bibitem{Hosseinidehaj_2015}
\BIBentryALTinterwordspacing
N.~Hosseinidehaj and R.~Malaney, ``Entanglement generation via non-gaussian
  transfer over atmospheric fading channels,'' \emph{Physical Review A},
  vol.~92, no.~6, dec 2015. [Online]. Available:
  \url{https://doi.org/10.1103%2Fphysreva.92.062336}
\BIBentrySTDinterwordspacing

\bibitem{Hosseinidehaj_2016}
\BIBentryALTinterwordspacing
------, ``{CV}-{QKD} with gaussian and non-gaussian entangled states over
  satellite-based channels,'' in \emph{2016 {IEEE} Global Communications
  Conference ({GLOBECOM})}.\hskip 1em plus 0.5em minus 0.4em\relax {IEEE}, dec
  2016. [Online]. Available:
  \url{https://doi.org/10.1109%2Fglocom.2016.7841711}
\BIBentrySTDinterwordspacing

\bibitem{Liao_2018}
\BIBentryALTinterwordspacing
Q.~Liao, Y.~Guo, D.~Huang, P.~Huang, and G.~Zeng, ``Long-distance
  continuous-variable quantum key distribution using non-gaussian
  state-discrimination detection,'' \emph{New Journal of Physics}, vol.~20,
  no.~2, p. 023015, feb 2018. [Online]. Available:
  \url{https://doi.org/10.1088%2F1367-2630%2Faaa8c4}
\BIBentrySTDinterwordspacing

\bibitem{Bohmann_2016}
\BIBentryALTinterwordspacing
M.~Bohmann, A.~A. Semenov, J.~Sperling, and W.~Vogel, ``Gaussian entanglement
  in the turbulent atmosphere,'' \emph{Physical Review A}, vol.~94, no.~1, jul
  2016. [Online]. Available: \url{https://doi.org/10.1103%2Fphysreva.94.010302}
\BIBentrySTDinterwordspacing

\bibitem{Tony2022}
\BIBentryALTinterwordspacing
T.~J.~G. Apollaro, S.~Lorenzo, F.~Plastina, M.~Consiglio, and K.~Życzkowski,
  ``Quantum transfer of interacting qubits,'' \emph{New Journal of Physics},
  vol.~24, no.~8, p. 083025, aug 2022. [Online]. Available:
  \url{https://dx.doi.org/10.1088/1367-2630/ac86e7}
\BIBentrySTDinterwordspacing

\bibitem{Huang_2021}
\BIBentryALTinterwordspacing
Y.-T. Huang, J.-D. Lin, H.-Y. Ku, and Y.-N. Chen, ``Benchmarking quantum state
  transfer on quantum devices,'' \emph{Physical Review Research}, vol.~3,
  no.~2, apr 2021. [Online]. Available:
  \url{https://doi.org/10.1103%2Fphysrevresearch.3.023038}
\BIBentrySTDinterwordspacing

\bibitem{Almeida_2018}
\BIBentryALTinterwordspacing
G.~M. Almeida, F.~A. de~Moura, and M.~L. Lyra, ``Quantum-state transfer through
  long-range correlated disordered channels,'' \emph{Physics Letters A}, vol.
  382, no.~20, pp. 1335--1340, may 2018. [Online]. Available:
  \url{https://doi.org/10.1016%2Fj.physleta.2018.03.028}
\BIBentrySTDinterwordspacing

\bibitem{Hermes_2020}
\BIBentryALTinterwordspacing
S.~Hermes, T.~J.~G. Apollaro, S.~Paganelli, and T.~Macr{\`{\i} },
  ``Dimensionality-enhanced quantum state transfer in long-range-interacting
  spin systems,'' \emph{Physical Review A}, vol. 101, no.~5, may 2020.
  [Online]. Available: \url{https://doi.org/10.1103%2Fphysreva.101.053607}
\BIBentrySTDinterwordspacing

\bibitem{Shen_2023}
\BIBentryALTinterwordspacing
S.~Shen, C.~Yuan, Z.~Zhang, H.~Yu, R.~Zhang, C.~Yang, H.~Li, Z.~Wang, Y.~Wang,
  G.~Deng, H.~Song, L.~You, Y.~Fan, G.~Guo, and Q.~Zhou, ``Hertz-rate
  metropolitan quantum teleportation,'' \emph{Light: Science Applications},
  vol.~12, no.~1, may 2023. [Online]. Available:
  \url{https://doi.org/10.1038%2Fs41377-023-01158-7}
\BIBentrySTDinterwordspacing

\bibitem{Bo2022}
\BIBentryALTinterwordspacing
B.~Li, Y.~Cao, Y.-H. Li, W.-Q. Cai, W.-Y. Liu, J.-G. Ren, S.-K. Liao, H.-N. Wu,
  S.-L. Li, L.~Li, N.-L. Liu, C.-Y. Lu, J.~Yin, Y.-A. Chen, C.-Z. Peng, and
  J.-W. Pan, ``Quantum state transfer over 1200 km assisted by prior
  distributed entanglement,'' \emph{Phys. Rev. Lett.}, vol. 128, p. 170501, Apr
  2022. [Online]. Available:
  \url{https://link.aps.org/doi/10.1103/PhysRevLett.128.170501}
\BIBentrySTDinterwordspacing

\bibitem{Bohmann_2017}
\BIBentryALTinterwordspacing
M.~Bohmann, J.~Sperling, A.~A. Semenov, and W.~Vogel, ``Higher-order
  nonclassical effects in fluctuating-loss channels,'' \emph{Physical Review
  A}, vol.~95, no.~1, jan 2017. [Online]. Available:
  \url{https://doi.org/10.1103%2Fphysreva.95.012324}
\BIBentrySTDinterwordspacing

\bibitem{Vasylyev_2018}
\BIBentryALTinterwordspacing
D.~Vasylyev, W.~Vogel, and A.~A. Semenov, ``Theory of atmospheric quantum
  channels based on the law of total probability,'' \emph{Physical Review A},
  vol.~97, no.~6, jun 2018. [Online]. Available:
  \url{https://doi.org/10.1103%2Fphysreva.97.063852}
\BIBentrySTDinterwordspacing

\bibitem{Hofmann_2019}
\BIBentryALTinterwordspacing
K.~Hofmann, A.~A. Semenov, W.~Vogel, and M.~Bohmann, ``Quantum teleportation
  through atmospheric channels,'' \emph{Physica Scripta}, vol.~94, no.~12, p.
  125104, sep 2019. [Online]. Available:
  \url{https://doi.org/10.1088%2F1402-4896%2Fab36e0}
\BIBentrySTDinterwordspacing

\bibitem{Eduardo2021}
E.~Villaseñor, M.~He, Z.~Wang, R.~Malaney, and M.~Z. Win, ``Enhanced uplink
  quantum communication with satellites via downlink channels,'' \emph{IEEE
  Transactions on Quantum Engineering}, vol.~2, pp. 1--18, 2021.

\bibitem{Zhao2022MonteCS}
W.~Zhao, R.~Shi, X.~Ruan, Y.~Guo, Y.~Mao, and Y.~Feng, ``Monte carlo-based
  security analysis for multi-mode continuous-variable quantum key distribution
  over underwater channel,'' \emph{Quantum Information Processing}, vol.~21,
  2022.

\bibitem{Heim_2009}
\BIBentryALTinterwordspacing
B.~Heim, D.~Elser, T.~Bartley, M.~Sabuncu, C.~Wittmann, D.~Sych, C.~Marquardt,
  and G.~Leuchs, ``Atmospheric channel characteristics for quantum
  communication with continuous polarization variables,'' \emph{Applied Physics
  B}, vol.~98, no.~4, pp. 635--640, dec 2009. [Online]. Available:
  \url{https://doi.org/10.1007%2Fs00340-009-3838-8}
\BIBentrySTDinterwordspacing

\bibitem{Nedasada2017}
\BIBentryALTinterwordspacing
N.~H.~R. Malaney, ``Cv-mdi quantum key distribution via satellite,'' apr 2017.
  [Online]. Available: \url{https://doi.org/10.26421%2Fqic17.5-6}
\BIBentrySTDinterwordspacing

\bibitem{Keshav2023}
\BIBentryALTinterwordspacing
K.~Kasliwal, J.~P~N, A.~Jain, and R.~K. Bahl, ``Enhancing
  satellite‐to‐ground communication using quantum key distribution,''
  \emph{IET Quantum Communication}, vol.~4, no.~2, p. 57–69, mar 2023.
  [Online]. Available: \url{https://doi.org/10.1049/qtc2.12053}
\BIBentrySTDinterwordspacing

\bibitem{Wang_2020}
\BIBentryALTinterwordspacing
Z.~Wang, R.~Malaney, and B.~Burnett, ``Satellite-to-earth quantum key
  distribution via orbital angular momentum,'' \emph{Physical Review Applied},
  vol.~14, no.~6, dec 2020. [Online]. Available:
  \url{https://doi.org/10.1103%2Fphysrevapplied.14.064031}
\BIBentrySTDinterwordspacing

\bibitem{Khmelev_2021}
\BIBentryALTinterwordspacing
A.~V. Khmelev, A.~V. Duplinsky, V.~L. Kurochkin, and Y.~V. Kurochkin, ``Stellar
  calibration of the single-photon receiver for satellite-to-ground quantum key
  distribution,'' \emph{Journal of Physics: Conference Series}, vol. 2086,
  no.~1, pp. 1--5, dec 2021. [Online]. Available:
  \url{https://dx.doi.org/10.1088/1742-6596/2086/1/012137}
\BIBentrySTDinterwordspacing

\bibitem{Agnesi_2020}
\BIBentryALTinterwordspacing
C.~Agnesi, M.~Avesani, L.~Calderaro, A.~Stanco, G.~Foletto, M.~Zahidy,
  A.~Scriminich, F.~Vedovato, G.~Vallone, and P.~Villoresi, ``Simple quantum
  key distribution with qubit-based synchronization and a self-compensating
  polarization encoder,'' \emph{Optica}, vol.~7, no.~4, p. 284, apr 2020.
  [Online]. Available: \url{https://doi.org/10.1364%2Foptica.381013}
\BIBentrySTDinterwordspacing

\bibitem{Osborn2021}
\BIBentryALTinterwordspacing
J.~Osborn, M.~J. Townson, O.~J.~D. Farley, A.~Reeves, and R.~M. Calvo,
  ``Adaptive optics pre-compensated laser uplink to leo and geo,'' \emph{Opt.
  Express}, vol.~29, no.~4, pp. 6113--6132, Feb 2021. [Online]. Available:
  \url{https://opg.optica.org/oe/abstract.cfm?URI=oe-29-4-6113}
\BIBentrySTDinterwordspacing

\bibitem{wang2022exploiting}
X.~Wang, C.~Dong, J.~Wei, T.~Wu, T.~Li, H.~Yu, L.~Shi, and S.~Zhao,
  ``Exploiting potentialities for space-based quantum communication network:
  downlink quantum key distribution modelling and scheduling analysis,'' 2022.

\bibitem{Chin2019}
H.-M. Chin, D.~Zibar, N.~Jain, T.~Gehring, and U.~L. Andersen, ``Phase
  compensation for continuous variable quantum key distribution,'' in
  \emph{2019 Conference on Lasers and Electro-Optics (CLEO)}, 2019, pp. 1--2.

\bibitem{Lanning2021}
\BIBentryALTinterwordspacing
R.~N. Lanning, M.~A. Harris, D.~W. Oesch, M.~D. Oliker, and M.~T. Gruneisen,
  ``Quantum communication over atmospheric channels: A framework for optimizing
  wavelength and filtering,'' \emph{Phys. Rev. Appl.}, vol.~16, p. 044027, Oct
  2021. [Online]. Available:
  \url{https://link.aps.org/doi/10.1103/PhysRevApplied.16.04402}
\BIBentrySTDinterwordspacing

\bibitem{Zheng2023}
Z.~{Zheng}, Z.~{Chen}, L.~{Huang}, X.~{Wang}, and S.~{Yu}, ``{Performance
  analysis of quantum key distribution using polarized coherent-states in
  free-space channel},'' \emph{Chinese Physics B}, vol.~32, no.~3, p. 030306,
  Mar. 2023.

\bibitem{Takenaka_2017}
\BIBentryALTinterwordspacing
H.~Takenaka, A.~Carrasco-Casado, M.~Fujiwara, M.~Kitamura, M.~Sasaki, and
  M.~Toyoshima, ``Satellite-to-ground quantum-limited communication using a
  50-kg-class microsatellite,'' \emph{Nature Photonics}, vol.~11, no.~8, pp.
  502--508, jul 2017. [Online]. Available:
  \url{https://doi.org/10.1038%2Fnphoton.2017.107}
\BIBentrySTDinterwordspacing

\bibitem{Ying2018}
\BIBentryALTinterwordspacing
Y.~Guo, C.~Xie, P.~Huang, J.~Li, L.~Zhang, D.~Huang, and G.~Zeng,
  ``Channel-parameter estimation for satellite-to-submarine continuous-variable
  quantum key distribution,'' \emph{Phys. Rev. A}, vol.~97, p. 052326, May
  2018. [Online]. Available:
  \url{https://link.aps.org/doi/10.1103/PhysRevA.97.052326}
\BIBentrySTDinterwordspacing

\bibitem{Dequal_2021}
\BIBentryALTinterwordspacing
D.~Dequal, L.~T. Vidarte, V.~R. Rodriguez, G.~Vallone, P.~Villoresi,
  A.~Leverrier, and E.~Diamanti, ``Feasibility of satellite-to-ground
  continuous-variable quantum key distribution,'' \emph{npj Quantum
  Information}, vol.~7, no.~1, jan 2021. [Online]. Available:
  \url{https://doi.org/10.1038%2Fs41534-020-00336-4}
\BIBentrySTDinterwordspacing

\bibitem{Peng_2022}
\BIBentryALTinterwordspacing
Q.~Peng, Y.~Guo, Q.~Liao, and X.~Ruan, ``Satellite-to-submarine quantum
  communication based on measurement-device-independent continuous-variable
  quantum key distribution,'' vol.~21, no.~2, feb 2022. [Online]. Available:
  \url{https://doi.org/10.1007/s11128-022-03413-z}
\BIBentrySTDinterwordspacing

\bibitem{Chai2019}
\BIBentryALTinterwordspacing
G.~Chai, Z.~Cao, W.~Liu, S.~Wang, P.~Huang, and G.~Zeng, ``Parameter estimation
  of atmospheric continuous-variable quantum key distribution,'' \emph{Phys.
  Rev. A}, vol.~99, p. 032326, Mar 2019. [Online]. Available:
  \url{https://link.aps.org/doi/10.1103/PhysRevA.99.032326}
\BIBentrySTDinterwordspacing

\bibitem{Chai2020BlindCE}
G.~Chai, D.~Li, Z.~Cao, M.~Zhang, P.~Huang, and G.~Zeng, ``Blind channel
  estimation for continuous-variable quantum key distribution,'' \emph{Quantum
  Eng.}, vol.~2, 2020.

\bibitem{Guo2017}
Y.~Guo, C.~Xie, Q.~Liao, W.~Zhao, G.~Zeng, and D.~Huang,
  ``Entanglement-distillation attack on continuous-variable quantum key
  distribution in a turbulent atmospheric channel,'' \emph{Physical Review A},
  vol.~96, 2017.

\bibitem{Lu_2019}
\BIBentryALTinterwordspacing
H.~Lu, Z.-D. Li, X.-F. Yin, R.~Zhang, X.-X. Fang, L.~Li, N.-L. Liu, F.~Xu,
  Y.-A. Chen, and J.-W. Pan, ``Experimental quantum network coding,'' \emph{npj
  Quantum Information}, vol.~5, no.~1, oct 2019. [Online]. Available:
  \url{https://doi.org/10.1038%2Fs41534-019-0207-2}
\BIBentrySTDinterwordspacing

\bibitem{Pande_2022}
\BIBentryALTinterwordspacing
V.~R. Pande and S.~Kanjilal, ``Quantum information transfer using weak
  measurements and any non-product resource state,'' \emph{Quantum Information
  Processing}, vol.~21, no.~3, feb 2022. [Online]. Available:
  \url{https://doi.org/10.1007%2Fs11128-022-03448-2}
\BIBentrySTDinterwordspacing

\bibitem{Semenov_2010}
\BIBentryALTinterwordspacing
A.~A. Semenov and W.~Vogel, ``Entanglement transfer through the turbulent
  atmosphere,'' \emph{Physical Review A}, vol.~81, no.~2, feb 2010. [Online].
  Available: \url{https://doi.org/10.1103%2Fphysreva.81.023835}
\BIBentrySTDinterwordspacing

\bibitem{Chuu2021}
\BIBentryALTinterwordspacing
C.-S. Chuu, C.-Y. Cheng, C.-H. Wu, C.-Y. Wei, S.-Y. Huang, Y.-J. Chen, S.-W.
  Feng, and C.-Y. Yang, ``Purification of single and entangled photons by
  wavepacket shaping,'' \emph{Advanced Quantum Technologies}, vol.~4, no.~3, p.
  2000122, 2021. [Online]. Available:
  \url{https://onlinelibrary.wiley.com/doi/abs/10.1002/qute.202000122}
\BIBentrySTDinterwordspacing

\bibitem{Wu_2019}
\BIBentryALTinterwordspacing
C.-H. Wu, C.-K. Liu, Y.-C. Chen, and C.-S. Chuu, ``Revival of quantum
  interference by modulating the biphotons,'' \emph{Physical Review Letters},
  vol. 123, no.~14, sep 2019. [Online]. Available:
  \url{https://doi.org/10.1103%2Fphysrevlett.123.143601}
\BIBentrySTDinterwordspacing

\end{thebibliography}

\end{document}